\documentclass[letterpaper,twocolumn,10pt]{article}
\usepackage{usenix,epsfig,endnotes,amsmath,bm,graphicx}
\usepackage{algorithm,algorithmic,booktabs,multirow,subfig}
\begin{document}
\date{}
\title{\Large \bf A Certified Robust Watermark For Large Language Models}
\author{
{\rm Xianheng Feng}\\
Zhejiang University\\
fengxianheng@zju.edu.cn
\and
{\rm Jian Liu*}\\
Zhejiang University\\
jian.liu@zju.edu.cn
\and
{\rm Kui Ren}\\
Zhejiang University\\
kuiren@zju.edu.cn
\and
{\rm Chun Chen}\\
Zhejiang University\\
chenc@zju.edu.cn
}
\maketitle

\thispagestyle{empty}

\begin{figure*}[t!]
\centering
\setlength{\abovecaptionskip}{-0.4cm}
\includegraphics[width=0.90\textwidth]{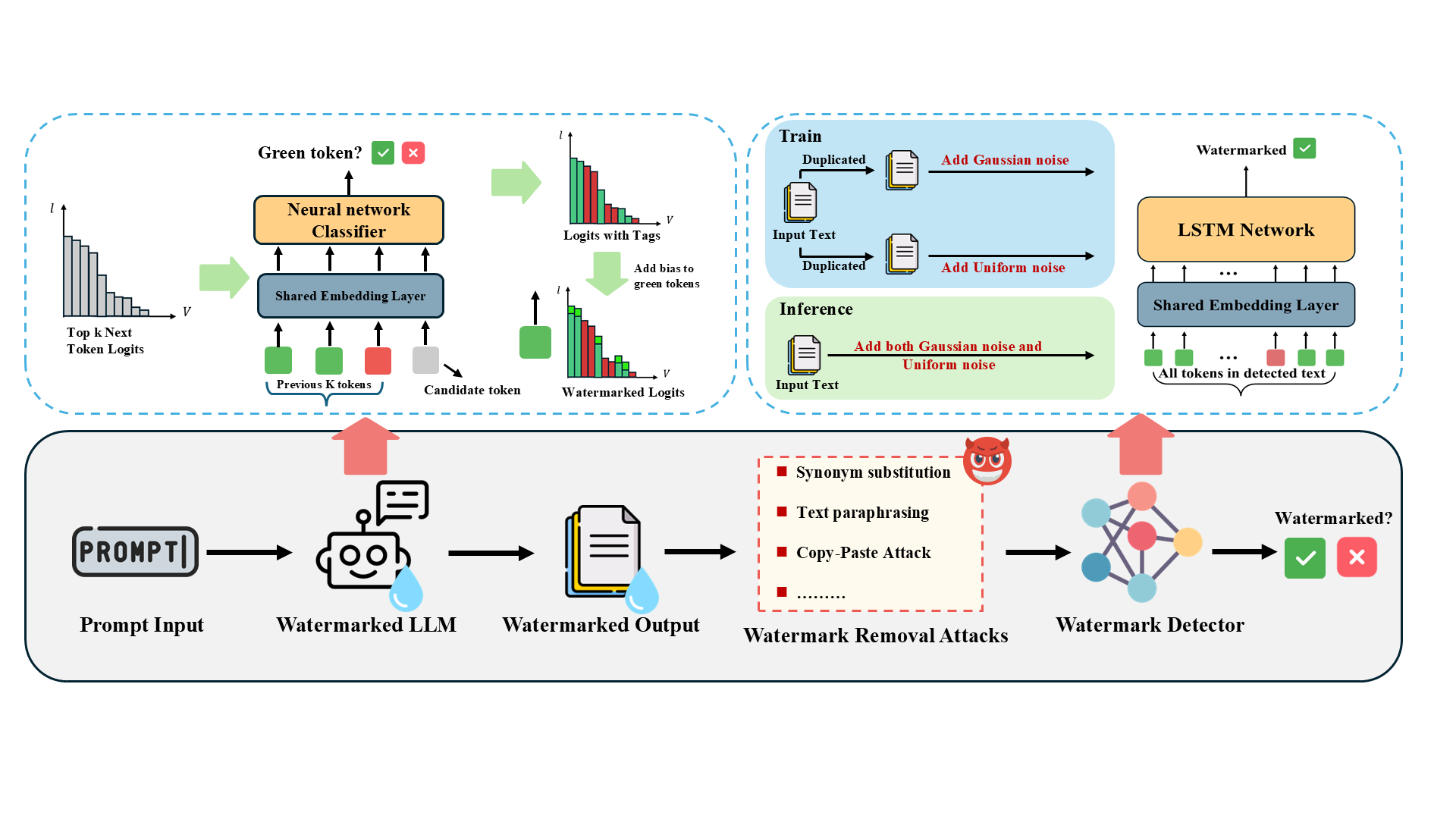} %
\caption{An overview of our certified robust watermarking algorithm. We utilize two different neural networks for watermark generation and detection. By adding Gaussian and Uniform noise during both training and inference stages, we improve the certified robustness of watermark algorithm and we are able to provide provable guarantees for watermarked text.}
\label{fig：overview}
\end{figure*}

\subsection*{Abstract}
The effectiveness of watermark algorithms in AI-generated text identification has garnered significant attention. Concurrently, an increasing number of watermark algorithms have been proposed to enhance the robustness against various watermark attacks. However, these watermark algorithms remain susceptible to adaptive or unseen attacks. To address this issue, to our best knowledge, we propose the first certified robust watermark algorithm for large language models based on randomized smoothing, which can provide provable guarantees for watermarked text. Specifically, we utilize two different models respectively for watermark generation and detection and  add Gaussian and Uniform noise respectively in the embedding and permutation space during the training and inference stages of the watermark detector to enhance the certified robustness of our watermark detector and derive certified radius. To evaluate the empirical robustness and certified robustness of our watermark algorithm, we conducted comprehensive experiments. The results indicate that our watermark algorithm shows comparable performance to baseline algorithms while our algorithm can derive substantial certified robustness, which means that our watermark can not be removed even under significant alterations.

\section{Introduction}
The rapid advancement of large language models (LLMs) such as ChatGPT\cite{chatgpt}, OPT\cite{LLM_OPT}, Claude\cite{claude} and LLAMA\cite{LLM_LLAMA} have brought transformative changes to respects of  our life for their high-quality response and the remarkable ability\cite{ablities1,ablities2,ablities3,ablities4} to generate human-like text. However, their out-standing capacity simultaneously presenting potential threats. Specifically, the high-quality text generated by LLMs can be also exploited by malicious individuals, leading to the dissemination of rumors and misinformation\cite{harmful_effect_fake_news,harmful_effect_misinformation}, which raises concerns of their potential misuse. Consequently, the detection and marking of text produced by LLMs are garnering increasing attention. 
\par
Watermarking is an effective solution for AI-generated text detection, which usually achieves AI-generated text identification by embedding specific patterns which are imperceptible to human's eyes but detectable by algorithms during text generation. For example, Kirchenbauer et al.\cite{kgw} proposes a watermark framework for large language models which works by randomly selecting green tokens form vocabulary according to the hash value of the previous k input tokens and slightly increasing the probability to be sampled of green tokens during text generation. However, this algorithm has been shown vulnerable to text adversarial attacks, especially text paraphrasing\cite{dipper}. To resolve this problem, Liu et al.\cite{SIR} proposes a semantic invariant robust watermark scheme, leveraging the semantic invariance of text before and after paraphrasing to mitigate the impact of text paraphrasing. And Zhao et al.\cite{unigram} suggests to use a fixed green token list during text generation to improve the robustness of watermark.
\par
However, while these methods effectively improve the robustness of watermark, they have their own limitations. Liu's algorithm requires knowledge of the user's input prompt, which is impractical in real-word scenarios. And Zhao's algorithm is weak at unforgeability, which refers to the difficulty of inferring watermarking rules from watermarked text. Moreover, these watermarking algorithms are robust only against specific empirical attacks, which means they may be vulnerable facing unseen attacks or adaptive attacks.
\par
A promising way to fight against unseen attacks is to provide provable robustness guarantee. In the fields of image and text classification, there has been a line of works\cite{certified_defense1,certified_defense2,li_RS,lecuyer_RS,cohen_RS,SAFER,CISS,text-crs} on certifiably robust defence to guarantee the robustness of model under unseen attacks. Among these method, randomized smoothing stands out as it imposes no constraints on the model's architecture and can be applied to large-scale models. Randomized smoothing smoothens the classifier by adding noise sampled from specific distribution into input during training stage and use a "smooth classifier"  constructed by trained classifier for prediction. It has been proven that one can easily obtain a guarantee that the classifier’s prediction is constant within some set around input $x$ by using randomized smoothing. 
\par
However, to our best knowledge, there is no certified robust watermark for LLMs based on randomized smoothing. To align with randomized smoothing, we propose the first certified robust watermarking algorithm for LLMs against  both mainstream watermark removal attacks and unseen attacks. In our work, following the framework of UPV\cite{UPV}, we use two different models with a shared embedding layer for watermark generation and detection and construct a certified robust watermark detector via randomized smoothing by adding Gaussian and Uniform noise respectively in text’s embedding and permutation space, where permutation space is an extra vector to indicate the alteration of token indices in watermark attacks, for example, the alteration from 'Who am I' to 'Who I am' can be viewed as a combination of no embedding alteration, because there is no change to the word/token set of the original sentence, but a permutation alteration between the second and third word/token indices. Moreover, we exhibit how to derive the robustness bounds in embedding and permutation space in our work.
\par
In the experiment, we evaluate certified accuracy and empirical robustness of our watermarking algorithm against various watermark attacks. Overall, our watermarking algorithm shows comparable and even superior performance compared to baseline on various watermark attacks and derives considerable certified radius on both embedding and permutation space, which means it is hard to erase our watermark even the watermarked text undergoes significant alterations.
\par
Our contributions are as follow:
\begin{itemize}
\item[$\bullet$]To our best knowledge, we propose the first certified robust watermarking algorithm.
\item[$\bullet$]We introduce randomized smoothing into watermarking and our watermark algorithm is certified robust under significant alterations. 
 \item[$\bullet$]We conducted extensive experiments to evaluate our algorithm. The experimental results demonstrate that our algorithm achieves comparable and even superior performance compared to baseline algorithms under various watermark attacks.
 \end{itemize}

\begin{table}[htbp] 
\caption{
\label{tab:notations}Summary of frequent notations} 
\setlength{\tabcolsep}{0.8mm}{
    \begin{tabular}{lcl} 
    \toprule Notation & Description \\ 
    \midrule $S$ & Sequence of tokens \\ 
    $M$ & Language model \\ 
    $G$ & Watermark generator \\
    $D$ & Watermark detector \\
    $D_{emb}$ & The embedding layers of watermark detector \\
    $D_{cls}$ & The classification layers of watermark detector \\
    $E$ & Actual embedding matrix \\
    $W$ & Embedding space \\
    $U$ & Permutation space \\
    $Y$ & Label space\\
    $\delta$ & Watermark strength\\
    $\phi(W,\,\varepsilon)$ & Embedding perturbation with parameter $W$ and  $\varepsilon$\\
    $\theta(U,\,\rho)$ & Permutation perturbation with parameter $U$ and  $\rho$\\
    \bottomrule 
    \end{tabular}
}
\end{table}

\section{PRELIMINARIES}
In this section, we introduce some necessary concepts to facilitate a better understanding of this paper.\\
\subsection{Watermark Algorithm} 
A complete watermark algorithm consists of two parts: watermark generation and watermark detection. In our work, following the framework of UPV\cite{UPV}, these two parts are carried out by two different neural networks: the watermark generator \emph{$G$} and watermark detector \emph{$D$}, respectively.
\begin{itemize}
\item[$\bullet$]\textbf{Watermark Gernation} : During text generation, a generative language model, denoted by \emph{$M$} , takes a sequence of tokens $S = [t_0,..., t_n-1]$ as prompt and generate logits vector $P = M(t_{0:n-1})$, which is a probability distribution over a vocabulary set \emph{$V$}. Meanwhile, watermark generator \emph{$G$} takes the last \emph{$K$} tokens of sequence \emph{$S$} as input and pick out the green tokens from vocabulary set \emph{$V$} . Increasing the logit value of green tokens by $\delta$, we can get a new logits vector $P' = G(P\,;\,t_{n-k:n-1})$ and the next token is selected from the new distribution \emph{$P'$} by either beam searching or sampling strategy.
\item[$\bullet$]\textbf{Watermark Detection} : Our watermark detector is a binary classification model. During watermark detection, our watermark detector takes an entire text as input and outputs the probability $P_w = D(S)$ that the text is watermarked.
\end{itemize}

\begin{figure}[t!]
        \centering
	\includegraphics[width=0.47\textwidth]{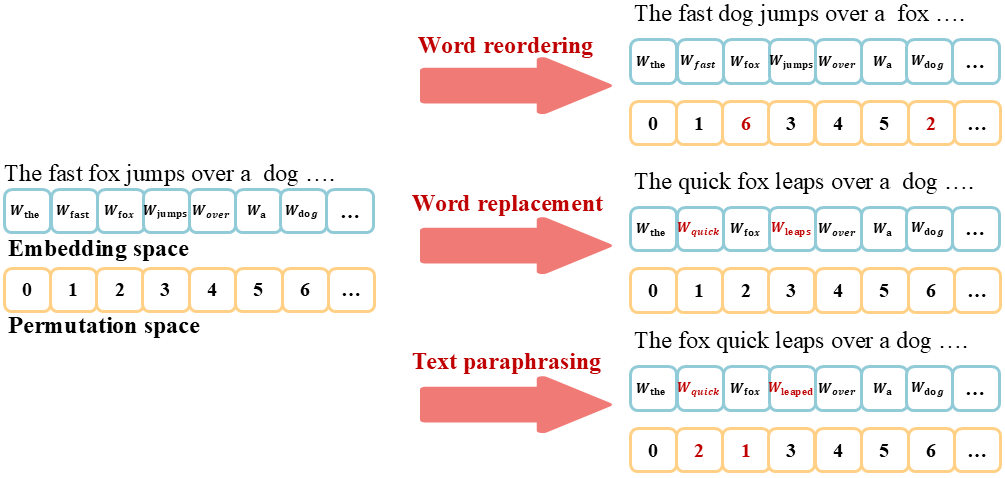} 
	\caption{The perturbation on embedding and permutation space under different text attacks.}
	\label{fig:permutation_example}
\end{figure}

\subsection{Watermark Attacks} 
Watermark attacks including watermark removal, watermark forgery, and watermark theft... are often used for malicious purposes.  Among these attacks, in this paper, we are more focused on watermark removal attacks which can be primarily divided into word-level attacks, such as synonym replacement, word deletion and word reordering, as well as sentence-level attacks that use language models to rewrite the entire text. In our paper, we regard text embedding $E = D_{emb}(S)$ as a combination of words embedding space $W =[w_1,...,w_n]$ and indexes permutation space $U =[u_1,...,u_n]$, where $D_{emb}$ is the  embedding layers of watermark detector (i.e. $E = W \bullet U$, where $\bullet$ means arranging the embedding vectors in $W$ according to vector $U$, to be concrete, each embedding $w_i$ vector should be located at the $ui$-th position of embedding matrix as shown in Figure \ref{fig:permutation_example}). Then watermark attacks can be viewed as adding perturbation to text's embedding and permutation space. Let  $\phi (W, \varepsilon)$ and  $\theta (U,\rho)$ be the perturbation function over embedding and permutation space by adding noise sampled from specific distribution and watermark attacks can be represented as $D_{emb}(A(S)) = \phi (W, \varepsilon) \bullet \theta (U,\rho)$, where $\varepsilon$ and $\rho$ are the noise sampled from specific distribution.

\subsection{Certifiably Robust Watermark Detector} 
 A classifier is certifiably robust means for any input x, one can  obtain a guarantee that the classifier’s prediction is constant within some set around x, which is often an $l_2$ or $l_{\infty}$ ball. And the classifier can be represented as followed:
\begin{equation}
    h(x) = h(x + \varepsilon) \quad \forall \, \Vert \varepsilon \Vert < R
\end{equation}

Where \emph{R} is called certified radius of input $S$. To get a certified classifier as mentioned, randomized smoothing may be a appropriate method. Applying randomized smoothing, we can construct a "smooth classifier" $g$ from arbitrary base classifier $h : R^d \xrightarrow{} Y$ that can be any deterministic or random function. When the input text 
x is given, the smooth classifier returns the class label whichever the base classifier is most likely to predict under the corruptions  of noise sampled from specific distribution \emph{F}:
\begin{equation}
    g(x) = \mathop{\arg\max}_{c \in Y}h(\varphi(S , \varepsilon)) 
\end{equation}
\qquad\qquad\qquad\qquad  where  \, $\varepsilon \sim F$\\
To construct a certifiably robust watermark detector, we apply randomized smoothing and obtain a smooth detector \emph{d}, which take text \emph{S} as input  and return the most likely label under specific corruptions:
\begin{equation}
    d(S)  = \mathop{\arg\max}_{c \in Y}(D_{cls}(\phi (W, \sigma) \bullet \theta (U,\rho))) 
\end{equation}
\qquad\qquad\qquad\qquad where $W \bullet U = D_{emb}(S)$ .

\section{PROPOSED METHOD}
In this section, we further introduces our method, the first certified robust watermark algorithm for large language model based on randomized smoothing. We firstly demonstrate the how we introduce randomized smoothing into watermark by adding noise to embedding and permutation space respectively along with the way to derive certified radius (Section \ref{subsection:randomized_smoothing}). Next, we explain the training stages of our method in detail and the improvement we adopt to achieve a more robust and effective watermarking algorithm(Section \ref{subsection:training_of_watermark_detector} to Section \ref{subsection:token_selection}). We finally present the complete algorithms for watermark generation and detection in Section \ref{subsection:the_framework}.

\subsection{Randomized Smoothing}
\label{subsection:randomized_smoothing}
To better describe the impact of various watermark attacks, as illustrated in Figure \ref{fig:permutation_example}, we introduce an additional permutation space to represent watermark attacks and regard attacks as the result of adding perturbation to both embedding and permutation spaces, which makes sense intuitively  because watermark attacks actually alter not only the composition of tokens but also tokens’ order in text. By introducing the permutation space, we separate the sequential attributes from the embedding matrix E, which allows us to consider watermark attacks from a finer-grained perspective. 
\par
Considering that Gaussian noise and Uniform noise can effectively simulate the impact on text’s content and order of  various watermark attacks such as synonym replacement, random deletion, and text paraphrasing, we introduce randomized smoothing to our watermark algorithm by respectively adding Gaussian noise and Uniform noise to the embedding and permutation spaces. In practice, we first divide the indices of tokens to groups and then randomly reorder them within groups to simulate adding Uniform noise to permutation space. The length of  each group is $\lambda$, which is the parameter of Uniform distribution. By utilizing existing theoretical tool, we are able to derive the certified radius on embedding and permutation space, which can be mathematically defined as follow:\\
$\mathbf{Theorem :}$ Let $\phi \,:\, \emph{W} \times R^{n \times d} {} \rightarrow{\emph{W}} $ be the perturbation function in embedding space based on Gaussian noise $\varepsilon \sim N(0,\sigma^{2}I)$ and $\theta \,:\, \emph{U} \times Z^{n} \rightarrow{\emph{U}}$ be the perturbation function in permutation space based on Uniform noise $\rho \sim U[-\lambda,\,\lambda]$. Let $d$ be the smooth detector from base detector $D$ as in (3) and suppose $y_A,y_B \in \mathbf{Y} $ and $ \underline{p_A},\overline{p_B} \in [0,1]$ $satisfy:$
\begin{equation}
\begin{aligned}
    P(D(\phi(W,\varepsilon)\bullet \theta(U,\rho)) = y_A) \,\geq\, \underline{p_A} \,\geq\, \overline{p_B} \,\geq\, \\
    \mathop{max}_{y_B \neq y_A}P(D(\phi(W,\varepsilon)\bullet \theta(U,\rho)) = y_B)
\end{aligned}
\end{equation}
then $d(\phi(W,\varepsilon_{0})\bullet \theta(U,\rho_{0})) \,=\, y_A$ for all $\Vert \varepsilon_{0} \Vert_2 < RAD_{e}$ and $\Vert \rho_{0} \Vert_1 < RAD_{p}$ , where $\varepsilon_{0}$ , $\rho_{0}$ is arbitrary noise on embedding and permutation space and
\begin{equation}
\begin{aligned}
    RAD_{e} \,=\, \sigma \Phi^{-1}(\underline{p_A})
\end{aligned}
\end{equation}
and 
\begin{equation}
\begin{aligned}
    RAD_{p} \,=\, \lambda(\underline{p_A} \,-\, \overline{p_B})
\end{aligned}
\end{equation}
\\
\emph{The proof of the theorem is provided in\cite{cohen_RS} and \cite{text-crs}}

\subsection{Training Stage of Watermark Detector}
\label{subsection:training_of_watermark_detector}
In the original framework of UPV, the watermark detector is trained using a randomly generated dataset where the green token ratio and tokens of each sample(i.e. the fraction of green tokens in text) are randomly selected. While using a randomly generated dataset rather than a dataset generated by one specific LLM enhances the detector's generalization ability, allowing the detector to detect watermarked text generated by various LLMs, it results in the z-score of each randomly generated sample being uniformly distributed between the lower and upper bound because the green ratio is uniformly sampled, which does not reflect the actual distribution of z-scores for non-watermarked and watermarked texts. Besides, this approach do not consider the frequency of tokens in real scenarios, for example, punctuation marks such as commas and periods may appear more frequently than other tokens in texts but they have the same frequency compared to other tokens in the original algorithm of UPV, which leads to a suboptimal training result 
especially when we apply randomized smoothing in training stage. To solve this issue, we use a LLM-generated watermark text dataset for detector’s training to migrate the difference between training dataset and test dataset .

We use LLM-generated dataset for training to reduce the difference of the z-score distribution and word frequency distribution between our training dataset and texts in real scenarios. As for the selection of LLM, we choose GPT-2\cite{gpt2}, a model with relatively high cross-entropy, for the generation of watermarked dataset. This is because models with lower cross-entropy such as LLaMA-7B is harder to get watermarked and may produce watermarked text with lower z-scores, sometimes even lower than non-watermarked samples, making it difficult for the detector to learn how to distinguish between watermarked and non-watermarked texts, thus degrading performance. However, since different LLMs use different tokenizers for encoding, the same word might be encoded into different tokens in different LLM, which means training with a dataset generated by a specific LLM may lead to poor generalization ability of our watermark detector. 

As a solution for this problem,  we utilize GloVe\cite{glove} tokenizer as a third-party tokenizer for input text encoding  instead of LLM tokenizers as in UPV, by which texts generated by different LLMs are encoded in the same way during detection. However, this solution also presents two challenges. The first challenge is that unlike LLM tokenizers which sometimes encodes long-term word to two or more tokens, GloVe tokenizer encodes each word to only one token . That means the number of tokens encoded by GloVe tokenizer corresponding to the same text is  fewer than LLM tokenizers, which is not conducive to watermark detection. We would solve this issue in Section \ref{subsection:encode_trick}. The second challenge is the inconsistency between watermark generator and watermark detector. While watermark detector uses GloVe tokenizer for encoding, our watermark generator should also select green tokens from GloVe’s vocabulary set rather than LLM tokenizer’s vocabulary set to keep consistency. To overcome this challenge, we modify the original green token selection scheme and it is discussed in Section \ref{subsection:token_selection}.

Additionally, to smoothen our detector and enhance its robustness, we introduce Gaussian and uniform noise into the training stage. We explored two different training strategies: one where both types of noise are simultaneously added to the input text during training, and another where the input text is duplicated, and each type of noise is added to one of the duplicates as input data for training. Surprisingly, the second method achieved better results than the first one and we present these results in Section \ref{section:experiment}.
\subsection{Encode Strategy} 
\label{subsection:encode_trick}
To ensure that the token sequence encoded by the GloVe tokenizer matches the length of token sequences encoded by original LLM tokenizer, we make some modifications to our encoding strategy. When the input text is encoded, we first use the LLM tokenizer to encode it into the corresponding token sequence. To convert tokens in that sequence into GloVe’s tokens without shortening the length of input token sequence, for each token at position \( i \) of token sequence, we decode the subsequence containing the previous \( N \) tokens (including the \( i \)-th token) into text using original LLM tokenizer and then encode it by GloVe tokenizer to get a new token sequence. We take the last token of this sequence as the converted token of  \( i \)-th token in original token sequence. By incorporating this encoding strategy, we ensure that the length of  token sequence encoded by GloVe tokenizer matches the original length of sequences encoded by LLM tokenizer. The complete process is as follows. 

\begin{algorithm}[!ht]
\floatname{algorithm}{Pseudocode}
\renewcommand\thealgorithm{}
\renewcommand{\algorithmicrequire}{\textbf{function}}
\renewcommand{\algorithmicensure}{\textbf{Output:}}
\caption{for text encoding}
\label{power}
\begin{algorithmic} 
\STATE \# GloVe tokenizer $\mathcal{G}$ and original tokenizer $\tau$
\STATE \# Token sequence $S = [t_{0},...,t_{n-1}]$ encoded by $\tau$ and Fixed integer $N$
\STATE $\bf{function}$ GloVeEncode$(S,\tau,\mathcal{G},N)$
\STATE \parbox{\linewidth}{\hangindent=1em \hangafter 0 1. Construct a token sequences List $L$, where each element $L_i$ corresponds to the token sequence $[t_{max(0,i-N)},...,t_{i}]$}
\STATE \parbox{\linewidth}{\hangindent=1em \hangafter 0 2. Decode elements in $L$ by $\tau$ and get text sequences List $L^{'}$}
\STATE \parbox{\linewidth}{\hangindent=1em \hangafter 0 3. Encode each text sequence in $L^{'}$ by $\mathcal{G}$ and get a new token sequences List $L^{''}$}
\STATE \parbox{\linewidth}{\hangindent=1em \hangafter 0 4. $\mathbf{return}$ token sequence $S^{'} = [L^{''}_0[-1],...,L^{''}_{n-1}[-1]]$ encoded by $\mathcal{G}$}

\end{algorithmic}
\end{algorithm}

\subsection{Green Token Selection}
\label{subsection:token_selection}
As in the original green tokens selection algorithm in UPV as shown in Algorithm \ref{original_Green_Token_Selection}, watermark generator select green tokens from candidate token set $T$ that contains tokens with the top $K$ logit value according to previous tokens which are encoded in LLM tokenizer’s way.  To keep the consistency between watermark generator and watermark detector, the input of watermark generator should be GloVe tokens. Therefore, we convert the input tokens to GloVe tokens as what we do in Section \ref{subsection:encode_trick} and modify the original algorithm as followed:

\begin{algorithm}[!ht]
\renewcommand{\algorithmicrequire}{\textbf{Input:}}
\renewcommand{\algorithmicensure}{\textbf{Output:}}
\renewcommand{\thealgorithm}{1}
\caption{Original Green Token Selection}
\label{original_Green_Token_Selection}
\begin{algorithmic}[1] 
\REQUIRE {Watermark Generator $G$. Previous token sequence $S = [t_0,...,t_{i-1}]$ and candidate token set $T$. Window size : \emph{w}. }
\ENSURE Green token set.
\FOR{candidate token $t$ in $T$}
\STATE Construct a token sequence $S = [t_{i-w},...,t_{i-1},t]$
\STATE Apply $S$ to wateramrk Generator $G$ and get the result $r = G(S)$. 
\IF{$r$ = 0}
\STATE Append t to Green token set.
\ENDIF
\ENDFOR
\STATE $\mathbf{return}$ Green token set.
\end{algorithmic}
\end{algorithm}

\begin{algorithm}[!ht]
\renewcommand{\algorithmicrequire}{\textbf{Input:}}
\renewcommand{\algorithmicensure}{\textbf{Output:}}
\renewcommand{\thealgorithm}{2}
\caption{Modified Green Token Selection}
\label{Modified_Green_Token_Selection}
\begin{algorithmic}[1] 
\REQUIRE {Watermark Generator $G$. LLM tokenizer $\tau$ and glove tokenizer $\mathcal{G}$. Previous token sequence $S = [t_0,...,t_{i-1}]$ and candidate token set $T$. Window size : \emph{w}. }
\ENSURE Green token set.
\FOR{candidate token $t$ in $T$}
\STATE Construct a token sequence $S = [t_{i-w},...,t_{i-1},t]$
\STATE Convert token sequence and get a new sequence $S^{'} = $ GloVeEncode$(S,\tau,g,N)$
\STATE Apply $S^{'}$ to wateramrk Generator $G$ and get the result $r = G(S^{'})$. 
\IF{$r$ = 0}
\STATE Append t to Green token set.
\ENDIF
 
\ENDFOR
\STATE $\mathbf{return}$ Green token set.
\end{algorithmic}
\end{algorithm}

\subsection{The Framework} 
\label{subsection:the_framework}
Combining the aforementioned content, the overall algorithm for generation and detection is described in detail in Algorithm \ref{Watermarked_Text_Generation} and \ref{Watermarked_Text_Detection}.

\begin{algorithm}[!ht]
\renewcommand{\algorithmicrequire}{\textbf{Input:}}
\renewcommand{\algorithmicensure}{\textbf{Output:}}
\renewcommand{\thealgorithm}{3}
\caption{Watermarked Text Generation}
\label{Watermarked_Text_Generation}
\begin{algorithmic}[1] 
\REQUIRE {Watermark Generator $G$. Fix Integer: K, N. Watermark strength: $\delta$. Language model: \emph{M}. Original LLM tokenizer $\tau$ and glove tokenizer $\mathcal{G}$. Prompt sequence: $S = [t_{-n},...,t_{-1}]$ encoded by $\tau$. Window size : \emph{w}. }

\FOR{$i = 0,1,...$}
\STATE Apply previous tokens $[t_{-n},...,t_{i-1}]$ to language model \emph{M} and compute the logits $P_{i} = M(t_{-n:i-1})$ of token $t_i$.

\STATE Extract top-K tokens from $P_{i}$, representing them as $T_i^{K} = TOPK(P_{i})$.

\STATE Construct a sequence matrix \emph{X}, where each row corresponds to the token sequence $X_i = [t_{i-N},...,t_{i}^j]$ for $t_{i}^j \in T_i^{K}$.

\STATE Transform tokens of each sequence in matrix to GloVe tokens and get new sequnce matrix $X^{'}$ where $X_j^{'} =$ GloVeEncode$(X_j,\tau,\mathcal{G},N)$.

\FOR{each sequence $X_j^{'}$ in $X^{'}$ }
\STATE Apply $X_j^{'}$ to watermark generator $G$ and get the output $r = G(X_j^{'})$.
\IF{$r = 1$}
\STATE Increase the logit value of $t_i^{j}$ by $\delta$.
\ENDIF
\ENDFOR
\STATE  Apply the softmax operator to the modified logits $P_{i}^{'}$ to get a probability distribution over vocabulary.

\STATE Predict $t_i$ token based on that probability distribution.
\ENDFOR
\end{algorithmic}
\end{algorithm}

During the actual watermark text detection phase, different from training, we need to add both Gaussian noise and Uniform noise to the text to obtain the certified radius for the embedding and permutation layers.

\begin{algorithm}[!ht]

\renewcommand{\algorithmicrequire}{\textbf{Input:}}
\renewcommand{\algorithmicensure}{\textbf{Output:}}
\renewcommand{\thealgorithm}{4}
\caption{Watermarked Text Detection}
\label{Watermarked_Text_Detection}
\begin{algorithmic}[1] 
\REQUIRE {Watermark Detector $D$. Fix Integer: $N$, $N_0$. Confidence parameter $\alpha$. Noise parameters $\sigma$ and $\lambda$. Original LLM tokenizer $\tau$ and GloVe tokenizer $\mathcal{G}$. Text sequence: $S = [t_{-n},...,t_{-1}]$ encoded by $\tau$.}

\STATE Transform $S$ to $S^{'} =$ GloVeEncode$(S,\tau,\mathcal{G},N)$ by encode trick.
\STATE Apply $S^{'}$ to detector's embedding layer and get embedding and permuation matrix $W \bullet U = E = D_{emb}(S^{'})$ .
\STATE Draw $N_0$ samples of noise, $\varepsilon_1,...,\varepsilon_{N_0} \sim N(0,\sigma^{2} I)$ and $N_0$ samples of noise, $\rho_1,...,\rho_{N_0} \sim U[-\lambda, \lambda]$ .

\STATE Apply each $E_i = (W+\varepsilon_i)\bullet (U+\rho_i)$ to the rest layers of detector.
\STATE Record the output of each input and get the counts of each class, denoting the most frequent class as $\hat{c_A}$ and the other class as $\hat{c_B}$, with their respective counts as $n_A$ and $n_B$
\STATE Compute the p-value of the two-sided hypothesis test that $n_A$ $\sim$ Binomial$(n_A + n_B; 0.5)$. 
\IF{p-value $\leq \alpha$}
\STATE $\mathbf{return}$ $\hat{c_A}$ and radius $RAD_e, RAD_p$ 
\ELSE
\STATE $\mathbf{return}$ $\mathbf{ABSTAIN}$
\ENDIF

\end{algorithmic}
\end{algorithm}




\section{Experiment}
\label{section:experiment}
\subsection{Experiment Setup}
\textbf{Dataset and Prompt.} Similar to UPV, we utilize the subset of  dbpedia\cite{dbpedia_c4} dataset and C4\cite{dbpedia_c4} dataset for data generation, taking the first 30 tokens as prompt and generate at least the next 200 tokens. Our training set consists of 5000 watermarked texts generated by GPT-2 and 5000 non-watermarked texts from dbpedia while our test set contains 500 generated watermarked texts and 500 non-watermarked texts from C4 dataset. \\
\textbf{Baseline Algorithm and Language Model.} We select two watermark algorithms as baseline method. One is Unigram\cite{unigram} an variant algorithm of KGW\cite{kgw} which use a fixed green token list rather than randomly divided green token list during next token generation. The other one is SIR\cite{SIR}, which utilize the semantic invariant feature of paraphrasing attack to improve the robustness of watermarking algorithm. To show the generalization of our method among different LLMs, we use GPT-2, OPT-1.3B, LLaMA-7B for data generation and evaluation. 
\\
\textbf{Evaluation.} To show the certified robustness of our method, we first evaluate the certified accuracy which is defined as the fraction of samples that is correctly classified and certified robust under different radii and explained how resilient our watermark algorithm is. Besides, we aslo compare our watermark with other algorithms by evaluating the F1-score of watermark detection under various empirical attacks. 
\\
\textbf{Hyper-parameters.}  The hyperparameters of our algorithm are configured as follows: window size 
$w$ of 2 ,watermark strength $\delta$ of 2  and K= 20 for watermark generator. For encoding, we  use the tokenizer of GloVe-6B as our GloVe tokenizer and set N = 30 to reduce time complexity. During the training stage of detector, we add Gaussian noise with $\sigma$ = 10,15,20,25 and $\lambda$ = 100,200 to evaluate detector’s performance under different perturbation and we optimize our detector by SGD\cite{sgd} with a learning rate of 0.1 while generator is trained via Adam\cite{adam} with a learning rate of 0.001.  Besides, we use a beam width of 3 for beam search strategy and set temperature = 0.7 for sampling strategy.
\subsection{Experiment Results}

\begin{figure}[!htb]
  \centering
  \subfloat[]{
    \includegraphics[scale=0.42]{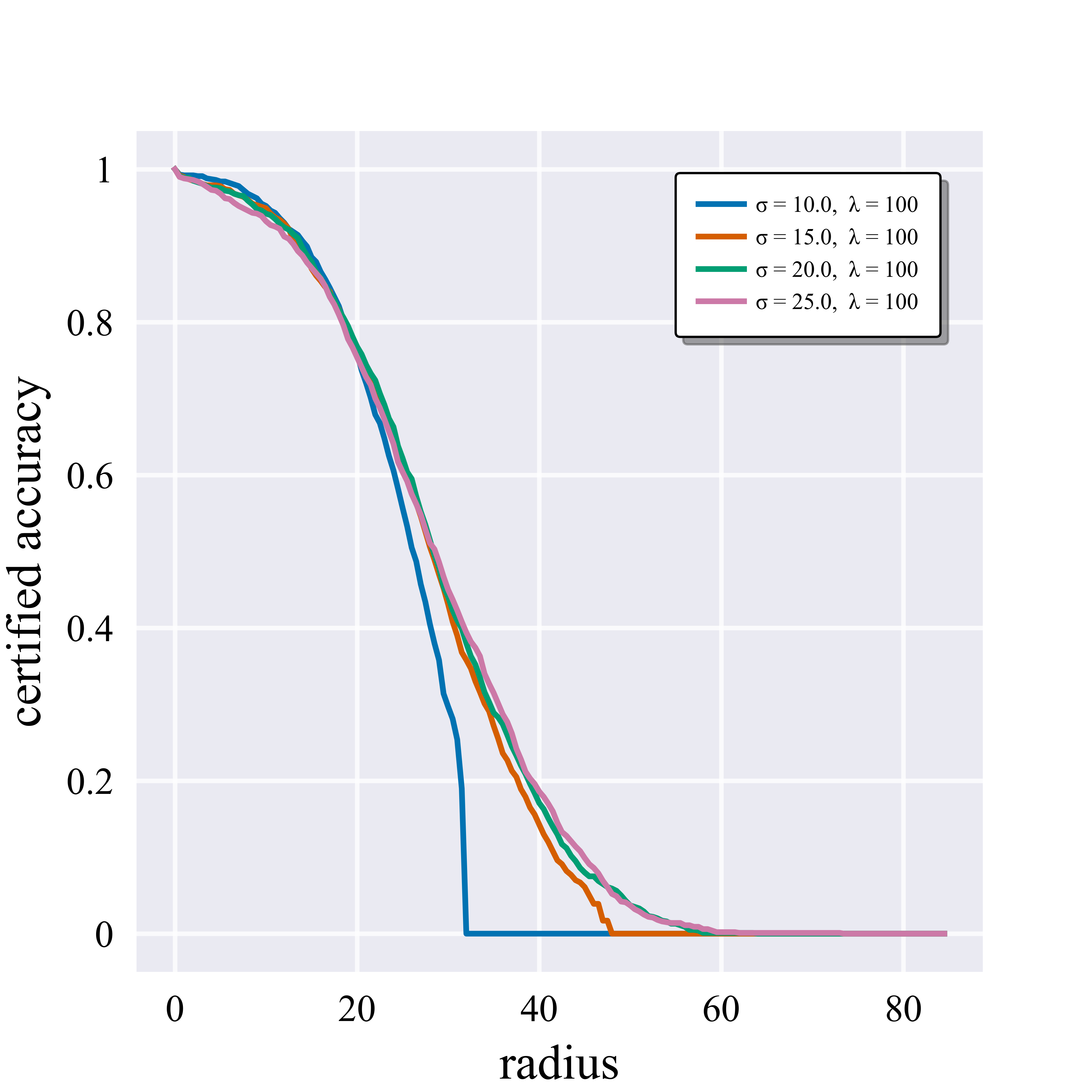}
  }
  \subfloat[]{
    \includegraphics[scale=0.42]{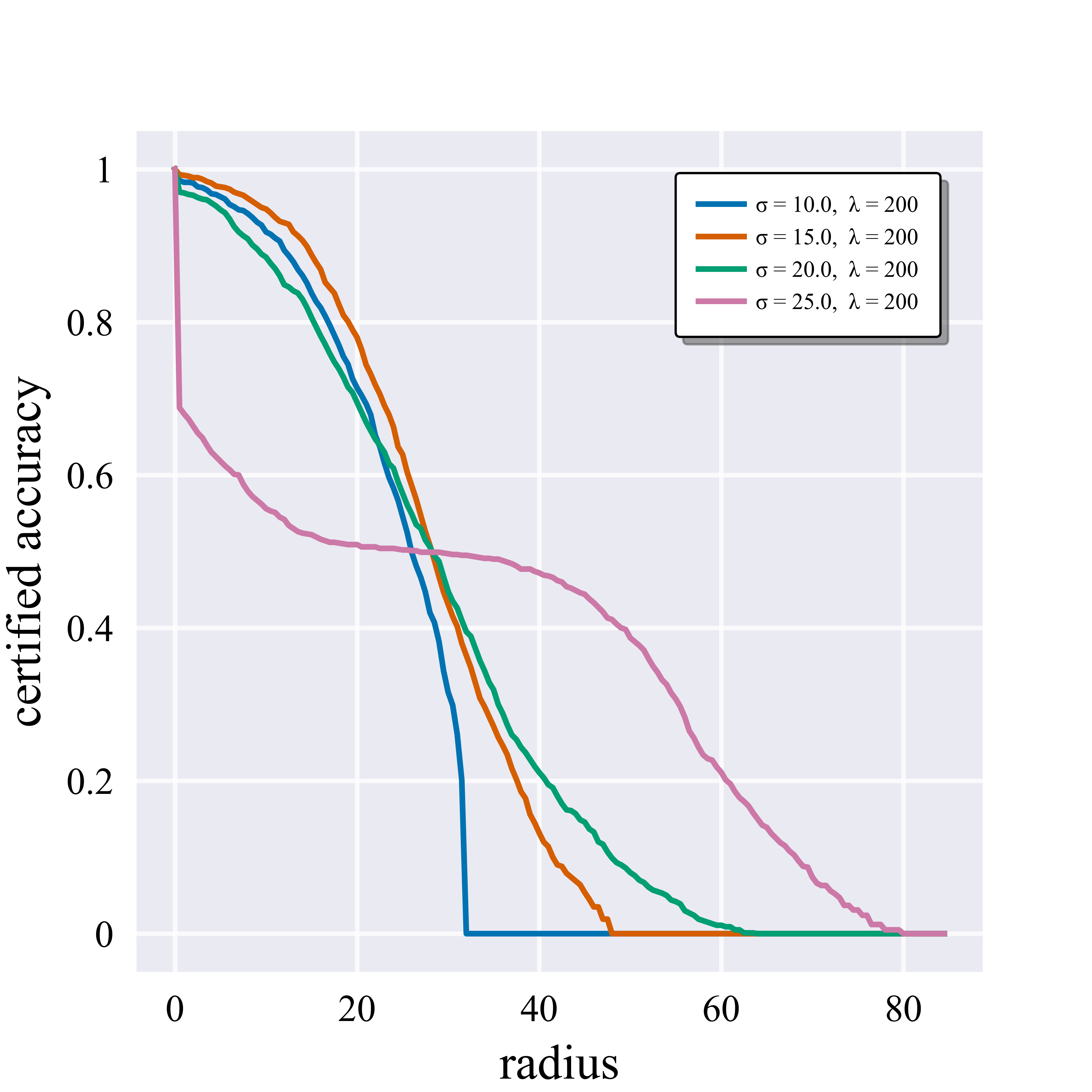}
  }
  \newline
  \subfloat[]{
    \includegraphics[scale=0.42]{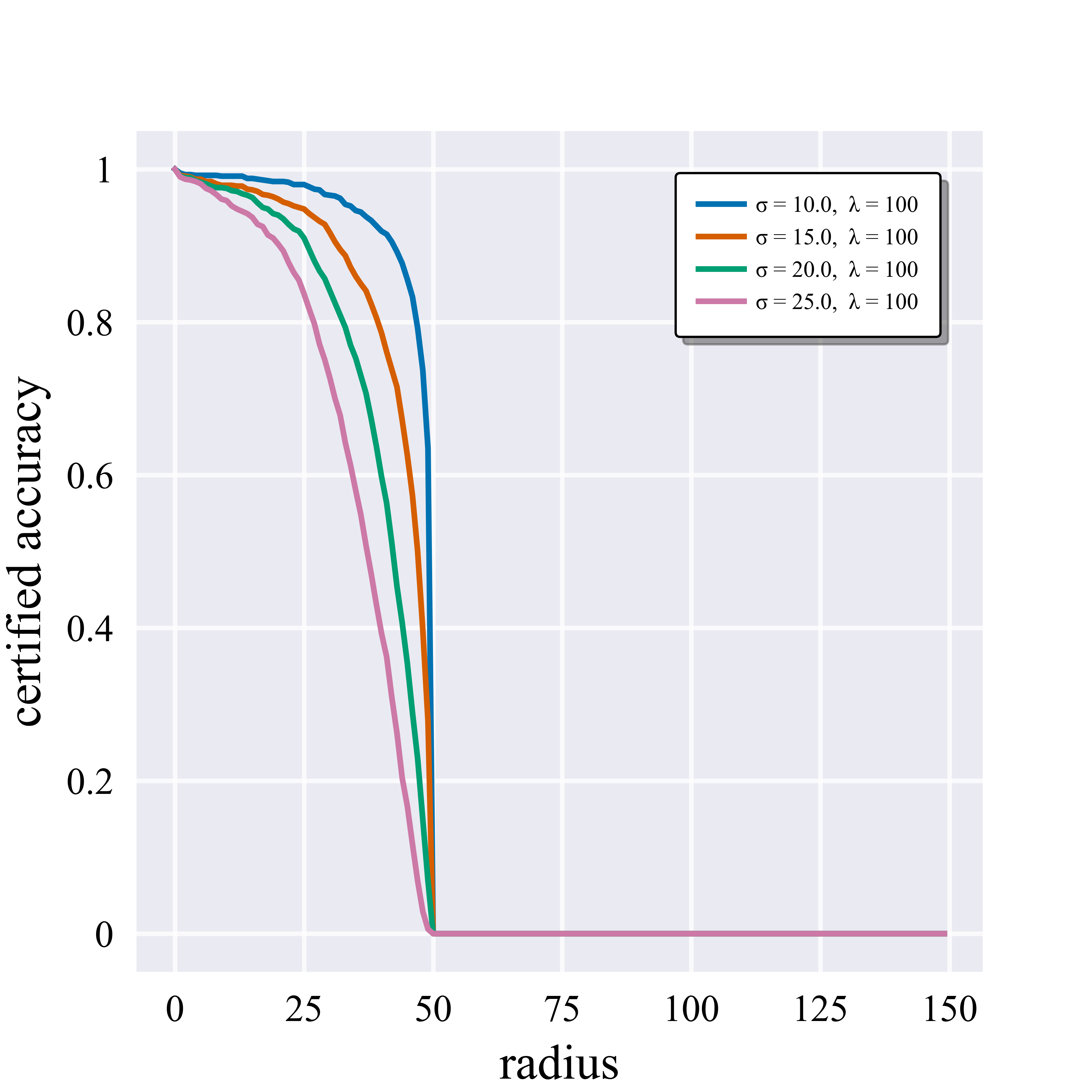}
  }
  \subfloat[]{
    \includegraphics[scale=0.42]{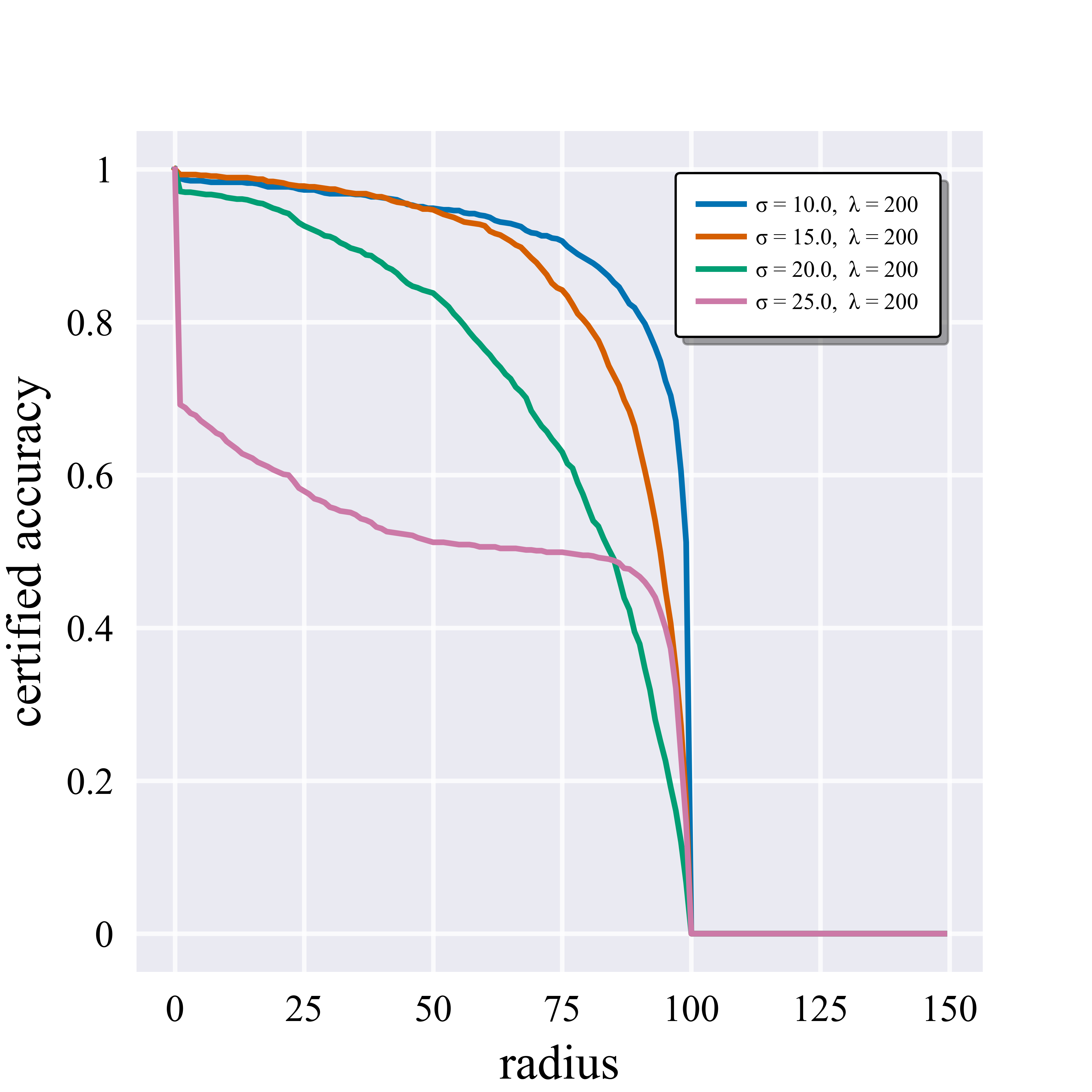}
  }
  \caption{Certified accuracy under different noise parameters setting.(a)(b) and (c)(d) are respectively the certify accuracy over embedding and permutation space.}
  \label{fig:certified accuracy}
\end{figure}

\subsubsection{Training Result}
To find a optimal noise setting, we train our watermark detector under combination of different levels of noise parameters $\sigma$ and $\lambda$.

Figure \ref{fig:certified accuracy}  exhibits the certified accuracy under different radii over embedding and permutation space of watermark detector trained with various level of noises. The higher the value of certify accuracy, the more certifiably robust the watermark is, and the stronger the guarantees it can provide. Figure \ref{fig:certified accuracy}(a) and Figure \ref{fig:certified accuracy}(b) illustrate the result that as we increase the intensity of the noise added to embedding space, the certify accuracy get improved in regions with higher radii without obvious decrease in regions with lower radii. However, when the variance of the noise exceeds a certain threshold, it either no longer enhances certify accuracy or leads to a sharp decline in certify accuracy in regions with lower radii. We also notice that the increment of $\lambda$ does not greatly effect the certified accuracy over embedding space, while Figure \ref{fig:certified accuracy}(c) and Figure \ref{fig:certified accuracy}(d) shows that as $\sigma$ increases, the certify accuracy over permutation space decreases consistently, which is particularly evident when $\lambda$ is set to 200.

\begin{figure}[t!]
        \centering
	\includegraphics[width=0.5\textwidth]{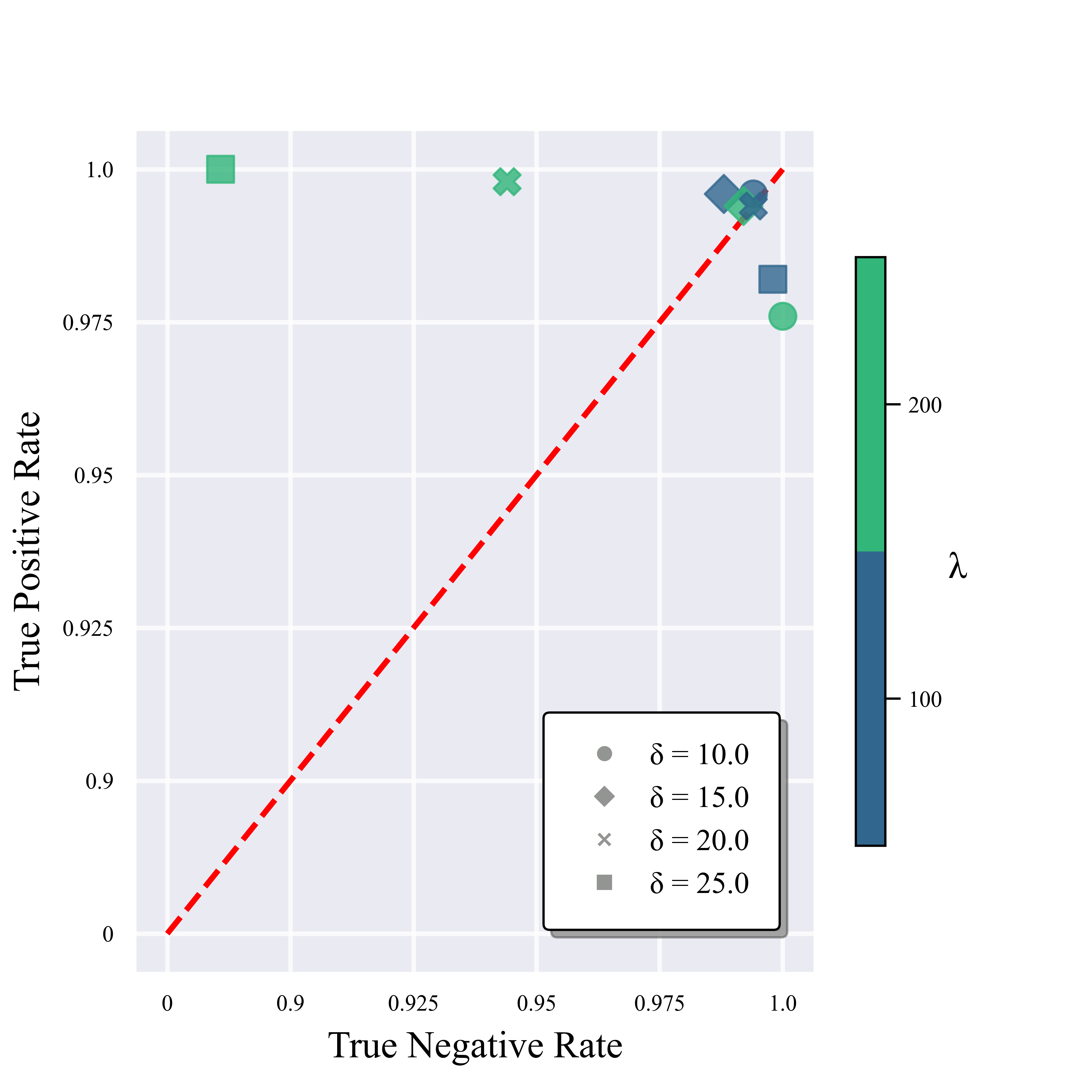} 
	\caption{The true positive rate and true negative rate under each combination of noise parameters.}
	\label{fig:fpr_tnr}
\end{figure}

In addition to certify accuracy, we also need to consider both the false positive rate and true positive rate of watermark detection. Figure \ref{fig:fpr_tnr} shows the true positive rate and true negative rate of each watermark detector over the test dataset generated by llama-7b using sampling strategy. We believe that a good watermark detector should maintain a high true positive rate while ensuring that the false positive rate is not excessively high. Points that deviate a lot from the red line in the figure are not ideal choices. After comprehensive evaluation, we consider $\sigma = 15.0$ and $\lambda = 200$ as the optimal noise parameters setting.

\subsubsection{Certified Robustness Analysis}
\label{subsection:Certified Robustness Analysis}
While we show the certified accuracy of each trained watermark detector, it is still ambiguous to find out how certifiably robust our watermark algorithm is and what extend of attack can our watermark tolerate.

To better understand the certified robustness of our watermark given the certified accuracy curve on embedding space, we analyzed the $l_2$ norm of embeddings of all tokens and the $l_2$ distance between themselves.

\begin{figure}[!htb]
  \centering
  \subfloat[]{
    \includegraphics[scale=0.26]{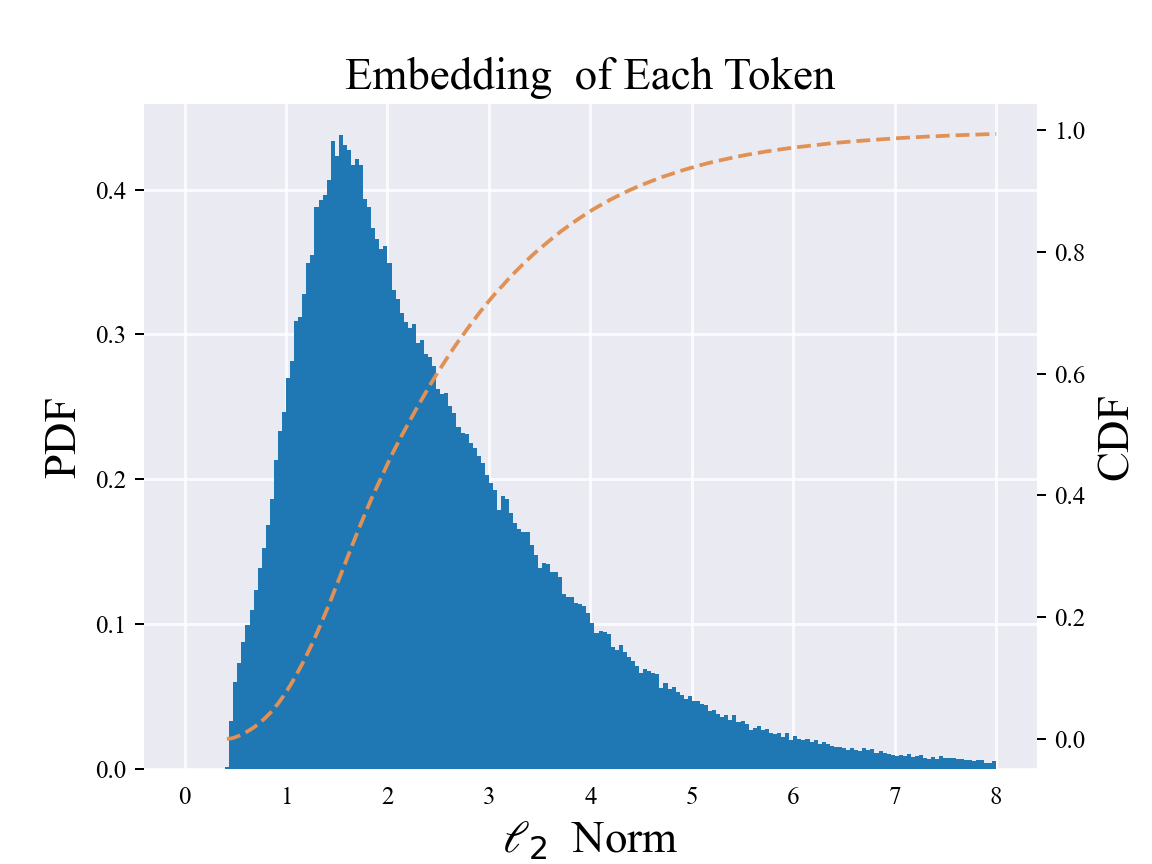}
  }
  \subfloat[]{
    \includegraphics[scale=0.26]{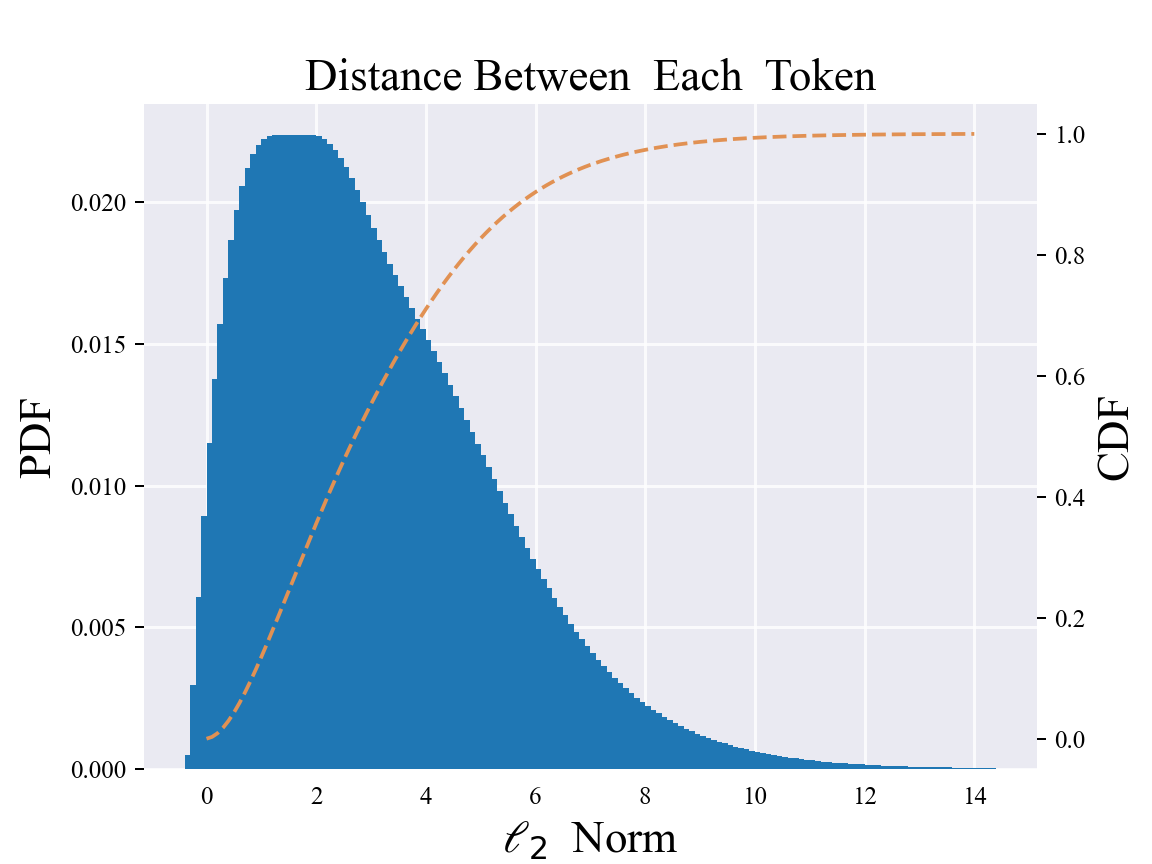}
  }
  \caption{(a) The PDF and CDF of embeddings' $l_2$ norm of all tokens. (b) The PDF and CDF of embeddings' $l_2$ norm distance between each token.}
  \label{fig:pdf_cdf}
\end{figure}

 As shown in Figure \ref{fig:pdf_cdf}, the $l_2$ norm of the embeddings is primarily concentrated between 1 and 3, while the $l_2$ distance is mainly distributed between 0 and 4. Through calculation, we obtained the average values of the $l_2$ norm of the embeddings and the $l_2$ distance as 2.5 and 3.2. Assuming that the certified radius over embedding space of a watermarked text is 20.0, then we can guarantee that the watermark will not be removed after about 44 tokens’ random replacement on average when the $l_2$  embedding distance between original text and altered text dose not exceed 20.0, which is actually considerable while about 80\% of certified radii are larger than 20.0 as shown in Figure \ref{fig:certified accuracy}. 

For the certified radius on permutation space, a larger radius means that our watermark is able to  keep certifiably robust  facing stronger attack perturbing the order of text. However, it is challenging to intuitively grasp the certified robustness of our method on permutation space from the values of the certified radius. Therefore, We compared the performance of watermark detectors respectively trained with and without adding noise on permutation space when watermarked texts are paraphrasing by DIPPER\cite{dipper} with two different settings.  For DIPPER-1, we set the lex diversity to 60 with order diversity of 0 and for DIPPER-2, we set its lex diversity the same as DIPPER-1 and increase its order diversity by 20, which means DIPPER-2 will make stronger perturbation on permutation space.

The result in Table \ref{tab:w/o uniform noise} shows that the watermark detector trained with noise performs better in all cases where the order of words is greatly shuffled and that training with Uniform noise improve the robustness of watermark.

\begin{table}[!htbp] 
\centering
\caption{
\label{tab:w/o uniform noise} The result of paraphrasing attack on the llama-7b test dataset using DIPPER-1 and DIPPER-2 when watermarked detector is trained with and without adding noise on permutation space and the variance of noise on embedding space is set to be 15.0.} 
\renewcommand{\arraystretch}{1.5}

\setlength{\tabcolsep}{2.0pt}{

    \begin{tabular}{cccccccc} 
    \toprule 
    \multirow{2}*{Attack} & \multirow{2}*{Method} & \multicolumn{3}{c}{Sampling} & \multicolumn{3}{c}{Beam Search}\\
    \cline{3-8}
    &&TPR&FPR&F1&TPR&FPR&F1\\
    \midrule
    
    \multirow{2}*{DIPPER-1} 
    &w/o noise&0.702&0.032&0.810&0.862&0.014&0.919\\
    &w/ noise &0.770&0.030&\textbf{0.856}&0.898&0.026&\textbf{0.933}\\
    \cline{1-8}
    \multirow{2}*{DIPPER-2} 
    &w/o noise&0.688&0.014&0.808&0.830&0.008&0.903\\
    &w/ noise &0.764&0.030&\textbf{0.851}&0.870&0.030&\textbf{0.916}\\
    \bottomrule 
    \end{tabular}
}

\end{table}

\begin{table*}[!htbp] 
\centering

\caption{\label{tab:paraphrasing attacks}The evaluation results of our watermark algorithm and baseline algorithms under various settings.} 
\renewcommand{\arraystretch}{0.96}
\setlength{\tabcolsep}{3mm}{
    \begin{tabular}{ccccccccc} 
    \toprule 
    \multirow{2}*{Setting} & \multirow{2}*{LLMs} & \multirow{2}*{Method} & \multicolumn{3}{c}{Sampling}  & \multicolumn{3}{c}{Beam Search}\\ 
    \cline{4-9}
    &&& TPR &FPR& F1  & TPR &FPR&F1\\
    \midrule
    \multirow{9}*{No Attack} 
    & \multirow{3}*{GPT-2} 
    &Unigram &1.000&0.012&\textbf{0.994}&1.000&0.012&\textbf{0.994}\\
    &&   SIR &0.996&0.012&0.992&1.000&0.012&\textbf{0.994}\\
    &&  Ours &0.994&0.012&0.991&0.996&0.012&0.992\\
    \cline{2-9}
    
    & \multirow{3}*{OPT} 
    &Unigram &1.000&0.012&\textbf{0.994}&1.000&0.012&\textbf{0.994}\\
    &&   SIR &0.995&0.012&0.992&1.000&0.012&\textbf{0.994}\\
    &&  Ours &0.990&0.012&0.989&0.992&0.012&0.990\\
    \cline{2-9}
    
    & \multirow{3}*{LLAMA} 
    &Unigram &0.992&0.008&0.992&0.998&0.012&\textbf{0.993}\\
    &&   SIR &0.996&0.008&\textbf{0.994}&0.996&0.012&0.992\\
    &&  Ours &0.994&0.008&0.993&0.998&0.012&\textbf{0.993}\\
    \cline{1-9}

    \multirow{9}*{DIPPER 1} 
    & \multirow{3}*{GPT-2} 
    &Unigram &0.888&0.024&0.929&0.952&0.030&0.960\\
    &&   SIR &0.701&0.024&0.813&0.830&0.030&0.892\\
    &&  Ours &0.936&0.024&\textbf{0.955}&0.962&0.030&\textbf{0.966}\\
    \cline{2-9}
    
    & \multirow{3}*{OPT} 
    &Unigram &0.615&0.024&0.751&0.862&0.034&0.909\\
    &&   SIR &0.689&0.024&0.805&0.876&0.034&0.917\\
    &&  Ours &0.834&0.024&\textbf{0.898}&0.944&0.034&\textbf{0.954}\\
    \cline{2-9}

    & \multirow{3}*{LLAMA} 
    &Unigram &0.687&0.030&0.800&0.886&0.026&0.927\\
    &&   SIR &0.794&0.030&\textbf{0.870}&0.898&0.026&\textbf{0.933}\\
    &&  Ours &0.770&0.030&0.856&0.898&0.026&\textbf{0.933}\\
    \cline{1-9}

    \multirow{9}*{DIPPER 2} 
    & \multirow{3}*{GPT-2} 
    &Unigram &0.960&0.038&\textbf{0.961}&0.968&0.040&\textbf{0.964}\\
    &&   SIR &0.810&0.038&0.876&0.820&0.040&0.882\\
    &&  Ours &0.940&0.038&0.950&0.962&0.040&0.961\\
    \cline{2-9}
    
    & \multirow{3}*{OPT} 
    &Unigram &0.645&0.048&0.762&0.846&0.054&0.890\\
    &&   SIR &0.635&0.048&0.755&0.836&0.054&0.884\\
    &&  Ours &0.768&0.048&\textbf{0.846}&0.924&0.054&\textbf{0.934}\\
    \cline{2-9}
    
    & \multirow{3}*{LLAMA} 
    &Unigram &0.631&0.030&0.760&0.890&0.030&\textbf{0.927}\\
    &&   SIR &0.725&0.030&0.826&0.874&0.030&0.918\\
    &&  Ours &0.764&0.030&\textbf{0.851}&0.870&0.030&0.916\\
    \cline{1-9}
    
    \multirow{9}*{GPT3.5} 
    & \multirow{3}*{GPT-2} 
    &Unigram &0.870&0.016&0.922&0.890&0.024&0.930\\
    &&   SIR &0.689&0.016&0.808&0.774&0.024&0.861\\
    &&  Ours &0.928&0.016&\textbf{0.956}&0.944&0.024&\textbf{0.959}\\
    \cline{2-9}
    
    & \multirow{3}*{OPT} 
    &Unigram &0.766&0.014&0.860&0.890&0.022&0.931\\
    &&   SIR &0.768&0.014&0.862&0.808&0.022&0.883\\
    &&  Ours &0.828&0.014&\textbf{0.899}&0.926&0.022&\textbf{0.951}\\
    \cline{2-9}
    
    & \multirow{3}*{LLAMA} 
    &Unigram &0.725&0.008&\textbf{0.837}&0.870&0.016&0.922\\
    &&   SIR &0.701&0.008&0.820&0.824&0.016&0.895\\
    &&  Ours &0.698&0.008&0.818&0.874&0.016&\textbf{0.925}\\
    
    \bottomrule 
    \end{tabular}
}
\end{table*}

\begin{table*}[!htbp] 
\centering
\caption{
\label{tab:other attacks}Performance of watermark algorithms under Emoji attack and Copy-Paste attack conducted using LLaMA-7b-chat and LLaMA-7b } 

\setlength{\tabcolsep}{9mm}{

    \begin{tabular}{ccccc} 
    \toprule 
    Attack & Method & TPR&FPR & F1\\
    \midrule
    \multirow{3}*{Emoji Attack} 
    &Unigram&0.888&0.010&0.936\\
    &SIR    &0.755&0.010&0.856\\
    &Ours   &0.870&0.002&0.929\\
    \cline{1-5}
    \multirow{3}*{Copy-Paste(150)} 
    & Unigram & 0.889&0.010 & 0.937 \\
    & SIR & 0.715&0.010 & 0.829 \\
    & Ours & 0.822&0.000 & 0.902 \\
    \cline{1-5}
    \multirow{3}*{Copy-Paste(200)} 
    & Unigram & 0.968&0.010& 0.979 \\
    & SIR & 0.912 &0.010&0.949 \\
    & Ours & 0.926 &0.004& 0.960 \\

    \bottomrule 
    \end{tabular}
}

\end{table*}

\begin{figure*}[t!]
\centering
    \subfloat[]{
    \includegraphics[scale=0.5]{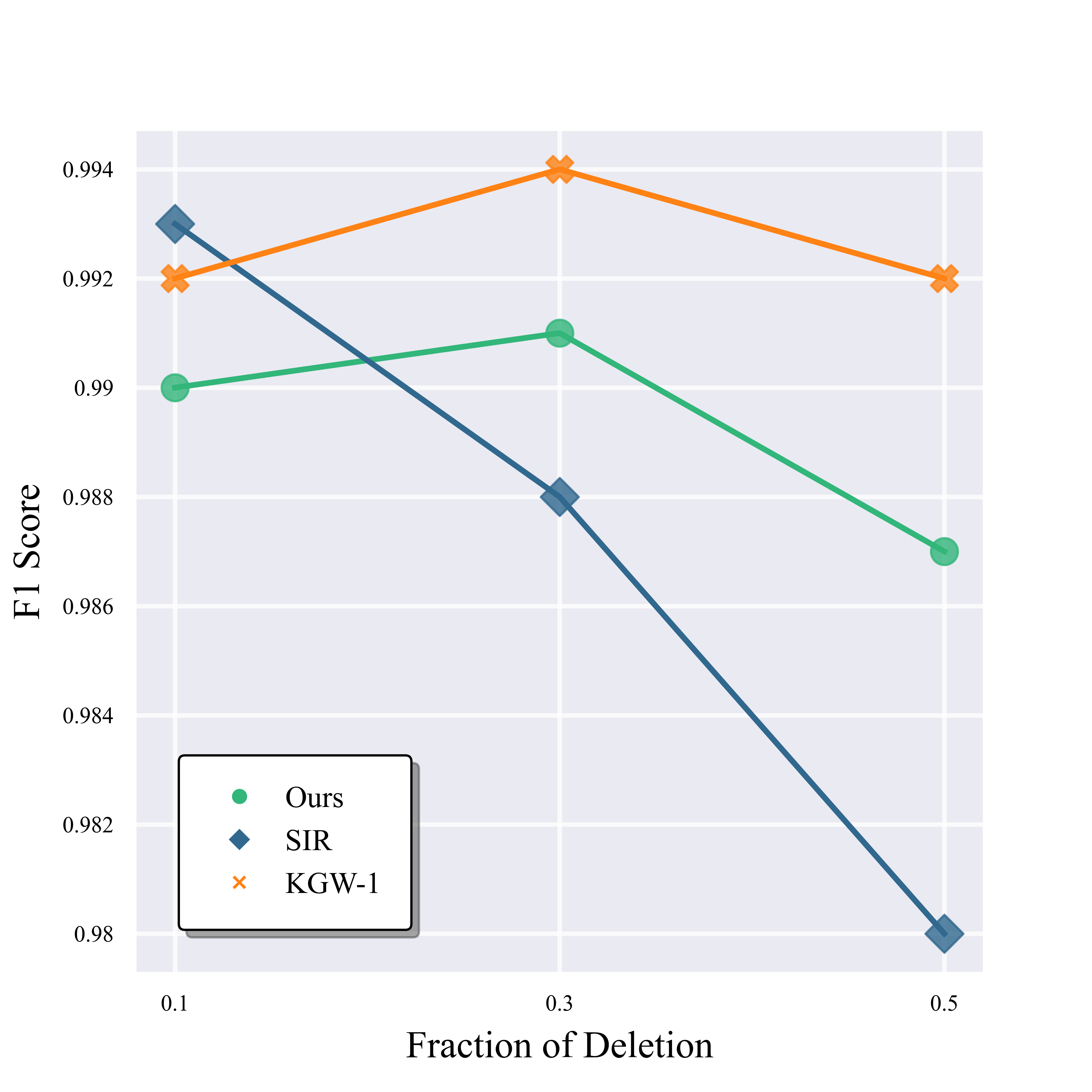}
    }
    \subfloat[]{
    \includegraphics[scale=0.5]{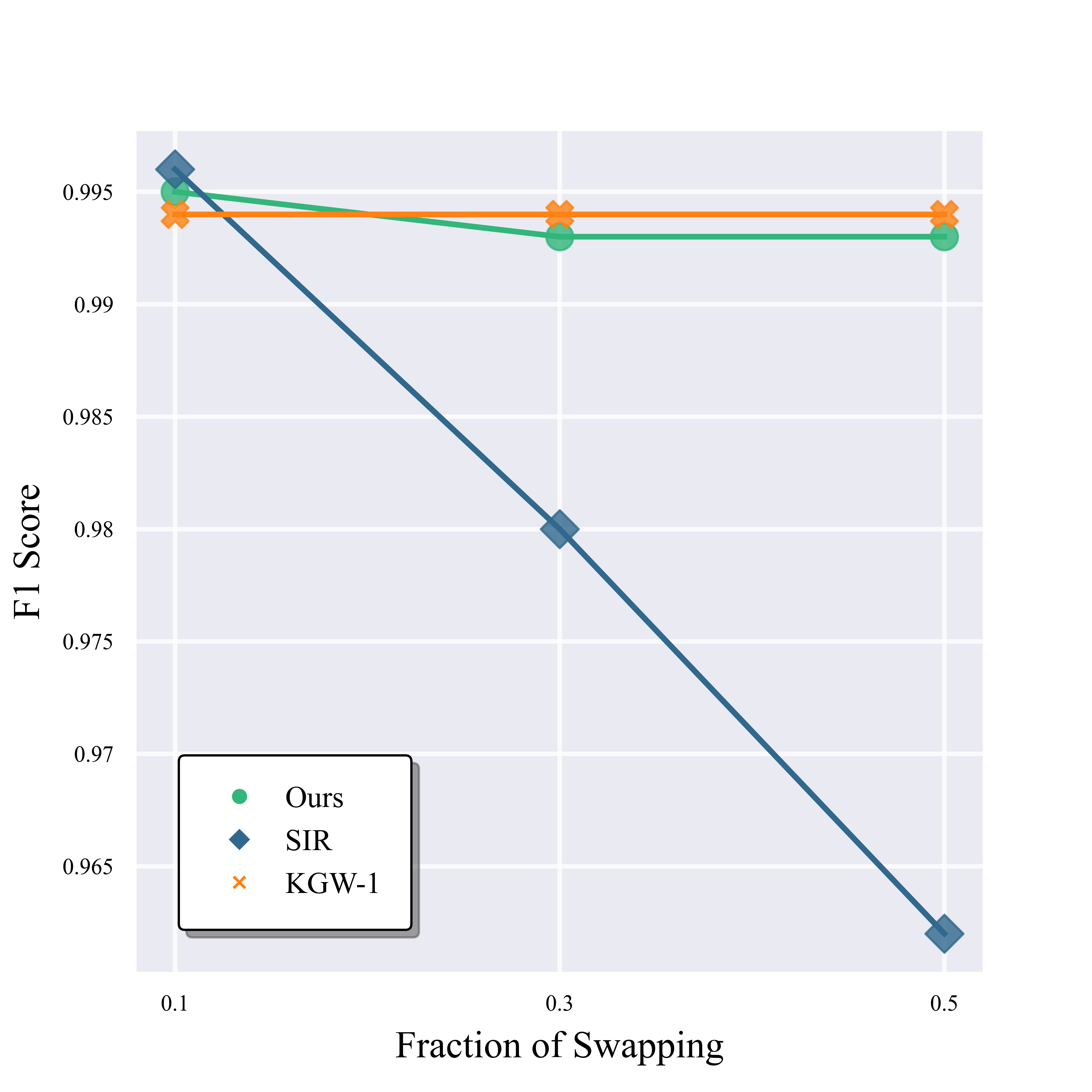}
    }
    \subfloat[]{
    \includegraphics[scale=0.5]{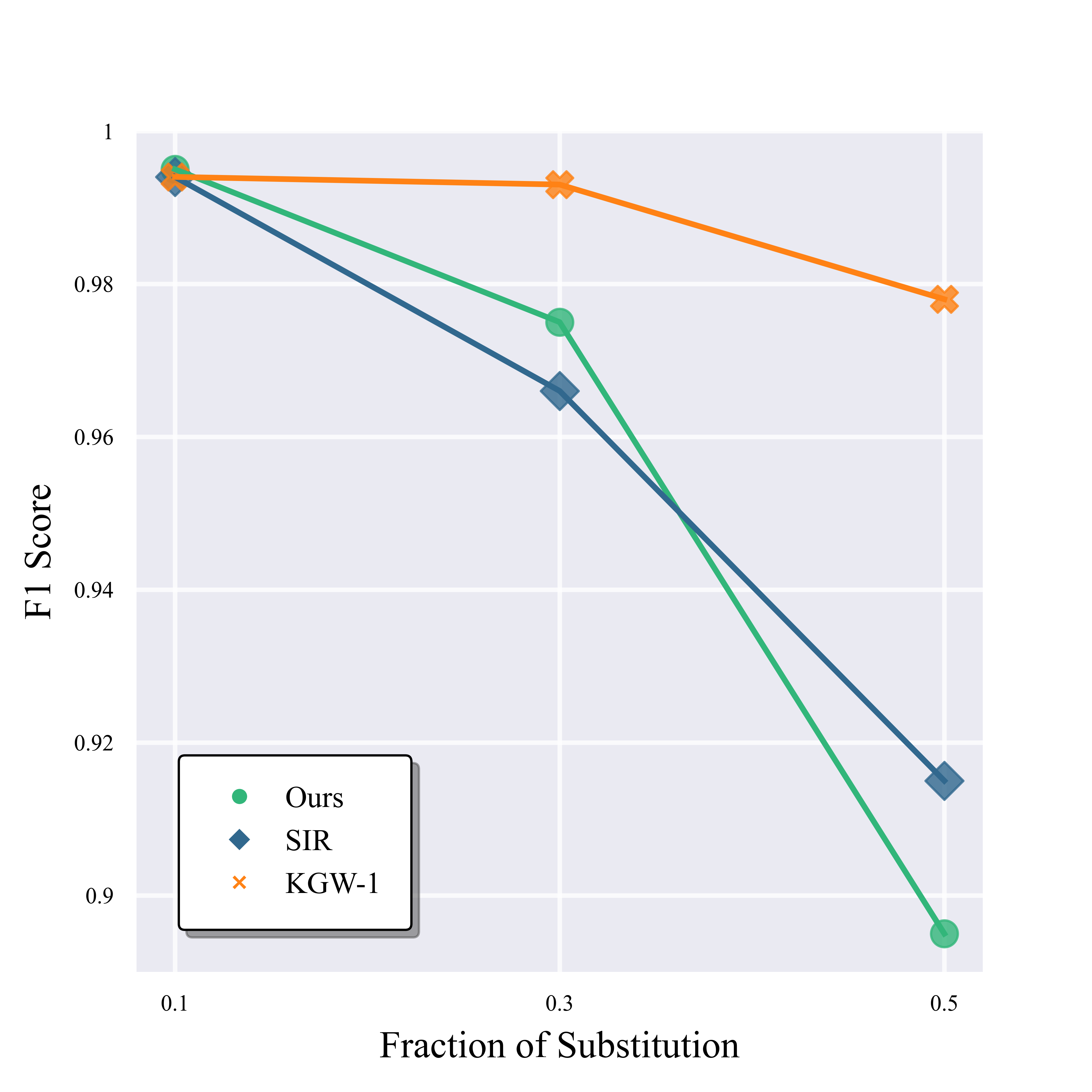}
    } 
\caption{This Figure show how the F1-score of three different watermark algorithms changes when the fraction of deletion, swapping and substitution attack increase.}
\label{fig:word_level_attack}
\end{figure*}

\subsubsection{Robustness against Empirical Attacks}
\textbf{Paraphrasing attacks.} Similar to SIR\cite{SIR}, we utilize several paraphrasing attacks including DIPPER-1, DIPPER-2 and ChatGPT to compare the robustness of baseline algorithms and our watermark algorithm under various settings. The settings of DIPPER-1 and DIPPER-2 are the same as mentioned in section \ref{subsection:Certified Robustness Analysis}.  For ChatGPT, we conduct attack using gpt-3.5-turbo-1103 using the prompt 'Rewrite the following paragraph:’ and we set the role of ChatGPT by telling it ‘You are a text paraphraser, your job is to rewrite texts’ to make sure it would follow the instruction. While SIR and Unigram can adjust their false positive rate by dynamically selecting a z-score threshold, the false positive rate of our algorithm can not be controlled to a fixed value. Therefore, for fair comparison, we set the false positive rate of SIR and Unigram to be the same as our algorithm and compare their F1-score. In addition, to validate the generalization ability of our watermark algorithm,  we also conduct experiment on various LLMs with different sizes including GPT-2, OPT-1.3B and LLAMA-7B. It is worth noting that in our experiments, unlike SIR, we paraphrase and detect the text excluding its prompt, which reflects real-world scenarios where exact prompt of a piece of AI-generated text is difficult to obtain.

Table \ref{tab:paraphrasing attacks} demonstrates that our algorithm outperforms Unigram and SIR in serveral cases when subjected to paraphrasing attacks while our algorithm also show comparable performance to baseline algorithms in other cases. For SIR, since we excludes the prompt during paraphrasing and detection in our experiment, it performs poorly when using sampling strategy. In no-attack setting, although our algorithm slightly underperforms compared to SIR and Unigram, which is due to the general feature of randomized smoothing, our algorithm still holds significant advantage which can not be overlooked that our algorithm is able to offers provable robustness guarantees for watermarked text. Moreover, our algorithm exhibits relatively stable performance across different LLMs, validating the generalizability of our algorithm. In contrast, SIR and Unigram have an obvious drop on their performance when employed in OPT-1.3B.

\textbf{Editing attacks.} In addition to the strongest paraphrasing attacks, we also evaluated the robustness of baseline algorithms and our algorithm against various editing attacks. These editing attacks represent relatively lower-cost and more common watermark removal attacks in practical scenarios. In this experiment, we fix the false positive rate of SIR and Unigram to 1\% and evaluate each algorithm’s F1-score on the test dataset generated using the beam search strategy with LLAMA-7B. The results, as shown in Figure \ref{fig:word_level_attack}, indicate that our method outperforms SIR but slightly underperforms compared to Unigram when the fraction increases in deletion and swapping scenarios. Moreover, although our method performs less robust than Unigram in synonym replacement scenarios, it still demonstrates comparable performance to SIR.

\textbf{Other attacks.} To make a comprehensive evaluation over our algorithm, in addition to paraphrasing attacks and editing attacks, we conduct Emoji and Copy-Paste attacks which are mentioned in \cite{kgw,reliability} to test the robustness of our algorithm in these two special scenarios. 

For the emoji attack, we adopt LLAMA-7B-Chat to generate watermarked text and prefix the instruction "Insert an asterisk * after each word." before each prompt.  After text generation, the asterisks would be removed from the text. As shown in Table \ref{tab:other attacks}, both our algorithm and Unigram demonstrated strong robustness against the emoji attack. Since the selection of green tokens in Unigram does not depend on the hash value of preceding token, the emoji attack has only a limited impact on it. Our algorithm, having incorporated token reordering noise during training, effectively reduces the impact of the emoji attack on the watermark.

For the Copy-Paste attack, we considered two concreate scenarios: inserting a piece of  watermarked text with 150 watermark tokens or 200 watermark tokens into a sequence of 600 human tokens. 

Considering that our detection model has a limit on input length and that a large number of unwatermarked tokens may significantly affect the model's predictions, we propose a sliding window detection approach to address this issue. For texts longer than 200 tokens, we used a window size of 200 tokens with a stride of 100 tokens to segment the entire text into several pieces. If any window segment is detected as watermarked by our watermark detector, the entire text was classified as watermarked. 

To ensure a fair evaluation, we also applied the same window segmentation scheme to the other two baseline algorithms in addition to their original detection methods. From two of the detection results, we select the best-performing result as the final performance under attack. For simplicity, when detecting with our algorithm, we only considered the worst-case scenario that might arise in the window segmentation context. For instance, in the case of inserting 150 watermark tokens, the worst case would be the case that a window containing 125 watermark tokens and 75 human tokens. Similarly, for the scenario of inserting 200 watermark tokens, the worst case would be a window containing 150 watermark tokens and 50 human tokens. Additionally, we randomly selected the human tokens to be inserted. The experimental results are shown in Table \ref{tab:other attacks}.

\subsubsection{The Impact of Text Length and Text Quality}
\begin{figure}[!htb]
  \centering
  \subfloat[]{
    \includegraphics[scale=0.42]{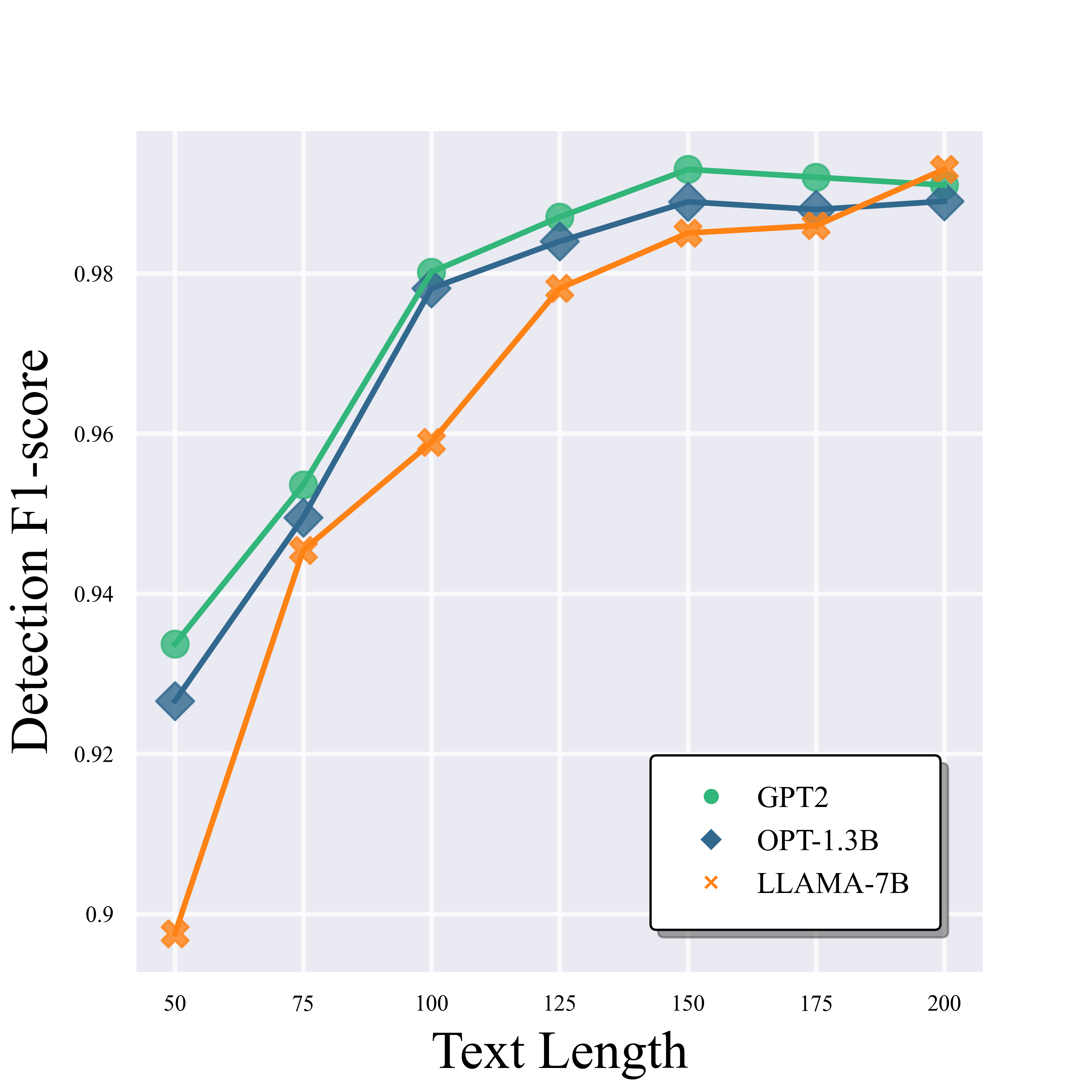}
  }
  \subfloat[]{
    \includegraphics[scale=0.42]{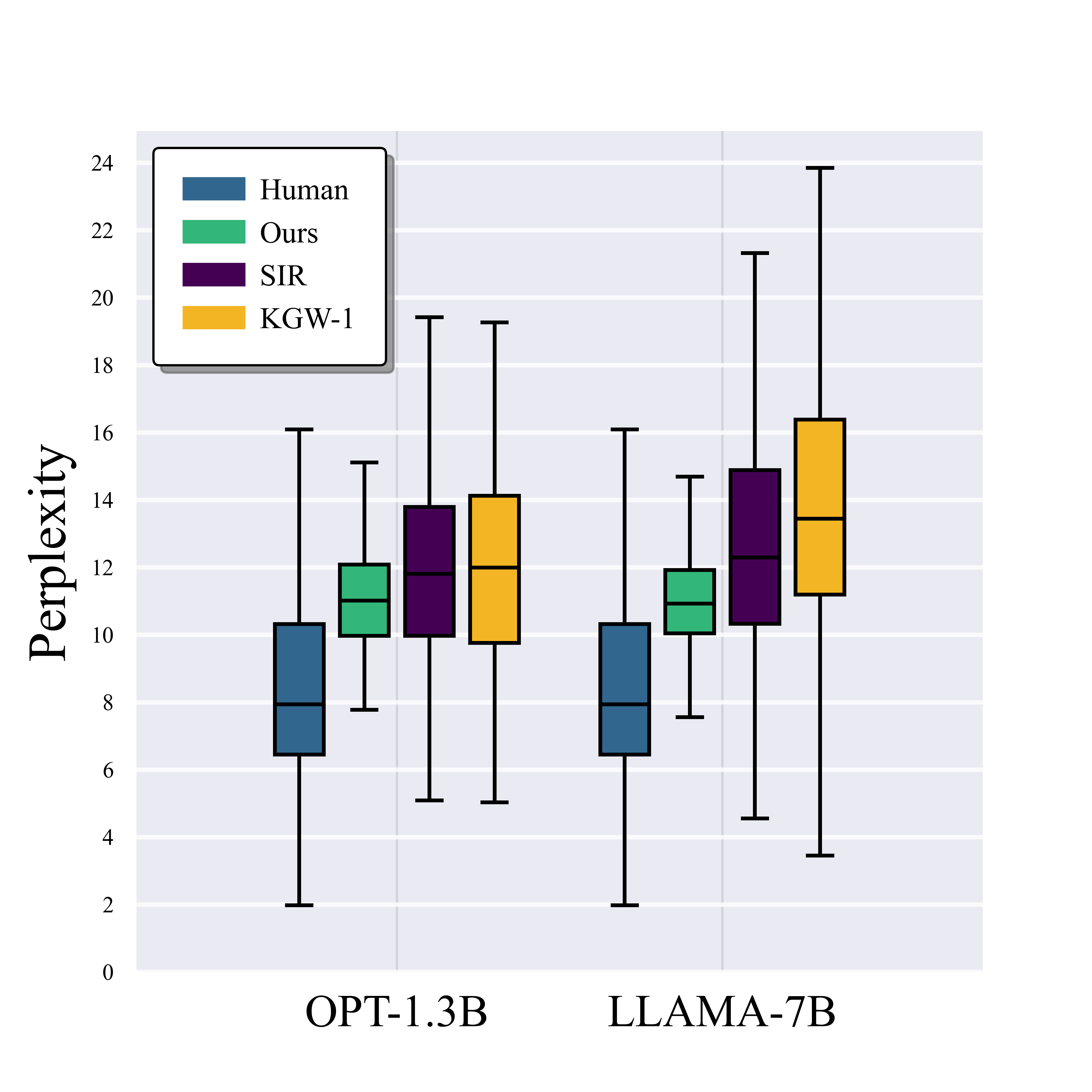}
  }
  \caption{(a) The impact of text generation length on detection F1-score of three different LLMs. (b) The perplexity of texts generated by different watermark algorithms.}
  \label{fig:length_impact_text_quality}
\end{figure}

\textbf{Impact of Text Length.} In figure \ref{fig:length_impact_text_quality}(a), we further evaluate the effectiveness of our watermark algorithm with different length of watermarked text generated by LLAMA-7B using beam search strategy. As shown, while the f1-score of watermark is far less than ideal when the length of text is 50, the score increase rapidly as text length increase from 50 to 125, with the score keeping stable as text length grow larger than 150. \\
\textbf{Text Quality.} To evaluation the influence of our watermark algorithm on text quality. We compare the perplexity of texts generated by different watermark algorithms in Figure \ref{fig:length_impact_text_quality}(b) and we utilize LLAMA-13B for the computation of perplexity. The texts to be evaluated are generated by OPT-1.3B and LLAMA-7B using beam search strategy. As shown in the Figure, while our watermark slightly increase the perplexity of texts compared to human-written texts, our watermark obtain lower perplexity than the other two baseline algorithms. Besides, the central region of our watermark is much smaller and concentrated, which indicates a stronger central tendency and reduced variability in the text generation using our watermark. 

\section{Related work}
\textbf{AI Generated Text Identification.} While large language models are applied to various scenarios with their increasingly powerful text generation capabilities, such as story generation\cite{story_generation}, article translation\cite{article_translation} and code assistance\cite{Code_Assistance}, the potential risks they bring\cite{harmful_effect_fake_news,harmful_effect_misinformation,harmful_effect_plagiarize} have made the identification of AI-generated text an urgent necessity. Currently, methods for identifying AI-generated text can be primarily divided into two categories: direct detection and watermarking.

Direct detection methods often identify AI-generated text by uncovering distinguishing characteristics between AI-generated and human-written texts or utilizing an extensive dataset of AI-generated text to train a accurate text classifier for AI-generated content. For example, regarding the first kind of method, Hamed and Wu\cite{bigrams_similarity} discovered that the bigram similarity of texts generated by ChatGPT is higher than that of human-written texts and developed a corresponding detection method based on this finding. Mitchell et al. found that AI-generated texts tend to occupy the negative curvature region of the model's log-probability function. Based on this observation they proposed DetectGPT\cite{detectGPT}. Building on Mitchell et al.'s approach, Bao et al.\cite{fast-detectGPT} improved the perturbation step in their method, significantly enhancing both its accuracy and speed. However, with the rapid advancement in both the size of model and text generation capabilities of LLMs, the gap between AI-generated texts and human-written texts has been narrowing\cite{essay,score_writing,comprehensiveLLM,abstract}. As a result, methods based on text characteristics are becoming increasingly less effective.For the second type of method, Mindner et al.\cite{multidimensional} employed multidimensional text feature extraction approaches to construct a classifier, with the best classifier outperforming GPTZero\cite{gptzero} in terms of F1-score. Chen et al.\cite{Roberta_T5} and Liu et al.\cite{Roberta} utilized the advanced language understanding capabilities of pretrained LLMs\cite{roberta_LLM,T5_LLM}, finetuning them as binary classifiers on various text datasets for AI-generated text detection. While these methods perform well on their respective test datasets, their effectiveness may be limited when applied to texts generated by newly emerging models. Likewise, as the capabilities of large language models continue to advance, their effectiveness remains a question.

Watermarking, as an alternative method for AI-generated text identification, is generally more effective, versatile, and interpretable compared to direct detection. Contemporary watermarking methods can be categorized based on the period of watermark insertion into two types: logits bias-based watermarking and token sampling-based watermarking. The first type involves biasing the logits of specific tokens to increase the probability of them to be sampled. Kirchenbauer et al.\cite{kgw} proposed a method where, during the generation of logits for each token, the hash values of the preceding k tokens are used to select the tokens as ‘green tokens’ for which the logit value should be increased by a fixed number. In their work, they also introduced a hypothesis testing method to determine whether a text is AI-generated, which achieve a outstanding detection accuracy with extremely low false positive rate upon AI-generated text. Based on this method, a line of works have been proposed to improve its performance in various aspects. Unigram\cite{unigram} propose to fix k to 1 to improve the robustness against watermark removal attacks while SIR\cite{SIR} thoroughly considered both the security robustness and attack robustness of watermark and propose to use a semantic invariant robust watermark model for green tokens selection. To improve the effectiveness of watermark in low-entropy scenarios,  EWD\cite{low_entropy_kgw} dynamically adjust the value to be added to logits according to tokens’ entropy value. Although that series of works have achieved great performance on AI-generated text identification, they reduce the quality of text for perturbing original text’s distribution. To address this issue, another kind of methods are proposed to add watermark by adjust the strategy of tokens sampling. Aaronson et al.\cite{Aaronson} applies the exp-minimum trick with a pseudo-random vector to reconstruct the process of sampling , by which they can detect whether the text is AI-generated through matching the patterns between texts and vector without reducing texts’ quality. Following this method, Kuditipudi et al.\cite{KTH} propose to use a pseudo-random sequence instead of a single pseudo-random vector and introduce  levenshtein distance during text detection,  which prevents LLMs provide the same outputs with the same prompt and improve robustness against watermark attacks.

In addition to these two primary identification methods mentioned above, some researchers have also combined features of both approaches. For example, Xu et al.\cite{xu_reinfore_watermark} first train a detector to detect the generated watermarked text and then tunes a LLM to generate text easily detectable by the detector while keeping its normal utility. Liu et al. propose an unforgeable publicly verifiable watermark algorithm called UPV\cite{UPV} where watermark generation and detection is processed by two different neural networks.

\textbf{Watermark Attacks.} In addition to character-level\cite{char_level_attack1,char_level_attack2} and word-level attack methods, new watermark attack techniques have emerged in response to the widespread use of large language model tools, targeting the flaws of watermarking algorithm and the unique application scenarios of AI-generated content. For instance, with strong capacity on text generation,  ChatGPT can be used to remove the watermark in text by paraphrasing. Regardless of ChatGPT, Krishna et al.\cite{dipper} also trains a specialized language model called DIPPER for text paraphrasing, which can rewrite and reorder text based on specified control conditions. Similar to rewriting attacks, the back-translation\cite{back_translation} method can also disrupt watermark by altering the content and structure of the texts. Besides, targeting watermarking algorithms that use the hash values of the previous k tokens to select “green tokens”, Kirchenbauer et al.\cite{kgw} proposed the "emoji attack," which instructs watermarked LLM to insert emojis between each word during text generation and then remove all emojis in text after generation to perturb the original hash values of previous k tokens. Gu et al.\cite{learnability} employed a watermarked LLM as teacher model to guide the training of a non-watermarked LLM student model, enabling the student model to directly produce watermarked texts and achieve watermark distillation. Furthermore, Sadasivan et al.\cite{human_误认为AI} discovered that even LLMs protected by watermarking schemes can be vulnerable to spoofing attacks aimed at misleading detectors into identifying human-written text as AI-generated, potentially causing reputational damage to developers.

\textbf{Randomized smoothing.} Randomized smoothing was originally applied in the field of image classification. Unlike other certified defense methods\cite{certified_defense1,certified_defense2,certified_defense3,certified_defense4,certified_defense5}, randomized smoothing can be used with large-scale models. Lecuyer et al.\cite{lecuyer_RS} first derive $l_2$ and $l_1$ robustness guarantee from randomized smoothing with Gaussian and Laplace noise and Li et al.\cite{li_RS}  subsequently improve the $l_2$ robustness guarantee for Gaussian noise. While all these guarantees are loose, Cohen et al.\cite{cohen_RS} provide a tight $l_2$ robustness guarantee for randomized smoothing. To develop models with certified robustness in NLP classification task, Ye et al. proposed SAFER\cite{SAFER}, a certified defense based on randomized smoothing which provides $l_0$ robustness guarantee against synonymous word substitution. Combining randomized smoothing and IBP\cite{IBP} method, CISS\cite{CISS} can provide $l_2$ robustness guarantee against word substitution attack in latent space. Although these method can provide provable robustness guarantee against substitution attack, they are not certifiably robust against other word-level attack such as deletion, swapping and insertion. To address this issue, Text-CRS\cite{text-crs} develop customized randomized smoothing theorems to derive certified robustness against these attacks.
\section{Conclusion} 

In this paper, we propose the first certified robust watermark algorithm. Specifically, we introduce randomized smoothing into watermark by adding Gaussian noise and Uniform noise respectively in embedding space and permutation space during the training of  watermark detector. To improve the performance of watermark detector while keep its generalization ability, we use real LLM-generated text for dataset and propose an encoding trick during tokenization. We conduct comprehensive evaluations of our watermark algorithm. The results show that our algorithm has comparable performance to baseline algorithms and even outperform in several cases while our watermark can provide considerable robustness guarantee.




\bibliographystyle{plain}
\bibliography{ref}

\end{document}